\documentclass [12 pt]{article}
\def\be{\begin{equation}}
\def\eea{\end{eqnarray}}
\def\bea{\begin{eqnarray}}
\def\ee{\end{equation}}

\usepackage[psamsfonts]{amssymb}
\usepackage{amsmath}
\author{F. Kheirandish$^{1}$ \footnote{fardin$_{-}$kh@phys.ui.ac.ir} and M.
Amooshahi$^{1}$ \footnote{amooshahi@sci.ui.ac.ir}
\\ $^{1}$ {\small Department of Physics, University of Isfahan,}
\\ {\small Hezar Jarib Ave., Isfahan, Iran.}}
\title{Minimal coupling method and the dissipative scalar field
theory}
\begin{document}
\maketitle
\begin{abstract}
\noindent Quantum field theory of a damped vibrating string as the
simplest dissipative scalar field investigated by its coupling
with an infinit number of Klein-Gordon fields as the environment
by introducing a minimal coupling method. Heisenberg equation
containing a dissipative term proportional to velocity obtained
for a special choice of coupling function and quantum dynamics for
such a dissipative system investigated. Some kinematical relations
calculated by tracing out the environment degrees of freedom. The
rate of energy flowing between the system and it's environment
obtained.
\end{abstract}
\section{Introduction}
In classical mechanics dissipation can be taken into account by
introducing a velocity dependent damping term into the equation
of motion. Such an approach is no longer possible in quantum
mechanics where a time-independent Hamiltonian implies energy
conservation and accordingly we can not find a unitary time
evolution operator for both states and observable quantities consistently.\\
To investigate the quantum mechanical description of dissipating
systems, there are some treatments, one can consider the
interaction between two systems via an irreversible energy flow
[1,2], or take a phenomenological treatment for a time dependent
Hamiltonian which describes damped oscillations, here we can
refer the interested reader to Caldirola-Kanai Hamiltonian for a
damped harmonic oscillator [3].
\begin{equation}\label{dm1}
H(t)=e^{-2\beta t}\frac{p^2}{2m}+e^{2\beta t}\frac{1}{2}m\omega^2
q^2
\end{equation}
There are significant difficulties about the quantum mechanical
solutions of the Caldirola-Kanai Hamiltonian, for example
quantizing with this Hamiltonian violates the uncertainty
relations or canonical commutation rules [4] and the uncertainty
relations vanish
as time tends to infinity and this is because the related time evolution operator is not unitary.\\
In 1931, Bateman [5] presented the mirror-image Hamiltonian which
consists of miror Hamiltonians, one of them represents the main
one-dimensional damped harmonic oscillator. Energy dissipated by
the main oscillator completely will be absorbed by the other
oscillator and thus the energy of the total system is conserved.
Bateman hamiltonian is given by
\begin{equation}\label{dm2}
H=\frac{p\bar{p}}{m}+\frac{\beta}{2m}(\bar{x}\bar{p}-xp)+(k-\frac{\beta^2}{4m})x\bar{x},
\end{equation}
with the corresponding Lagrangian
\begin{equation}\label{dm3}
L=m\dot{x}\dot{\bar{x}}+\frac{\beta}{2}(x\dot{\bar{x}}-\dot{x}\bar{x})-kx\bar{x},
\end{equation}
canonical momenta for this dual system can be obtained from this
Lagrangian as
\begin{equation}\label{dm3}
p=\frac{\partial L}{\partial
\dot{x}}=m\dot{\bar{x}}-\frac{\beta}{2}\bar{x},\hspace{1.50
cm\bar{p}}=\frac{\partial L}{\partial
\dot{\bar{x}}}=m\dot{x}+\frac{\beta}{2}x,
\end{equation}
dynamical variables $ x, p $ and $\bar{p},\bar{x} $ shoud satisfy
the commutation relations
\begin{equation}\label{dm4}
[x,p]=i,\hspace{2.00 cm} [\bar{x},\bar{p}]=i,
\end{equation}
however the time-dependent uncertainty products obtained in this
way, vanishes as time tends to infinity.\\
Caldirola [3,6] developed a generalized quantum theory of a linear
dissipative system in 1941 : equation of motion of a single
particle subjected to a generalized non conservative force $ Q $
can be written as
\begin{equation}\label{dm5}
\frac{d}{d t}(\frac{\partial T}{\partial \dot{q}})-\frac{\partial
T }{\partial q}=-\frac{\partial V}{\partial q}+Q(q),
\end{equation}
where $ Q_r =-\beta(t)\sum a_{r j}\dot{q}_j $, and $ a_{rj} $'s
are some constants, changing the variable $t$ to $t^{*}$ using the
following nonlinear transformation
\begin{equation}\label{dm6}
t^*=\chi(t),\hspace{1.50cm}dt=\phi(t)dt^*,\hspace{1.50cm}\phi(t)=e^{\int_0^t\beta(t')dt'},
\end{equation}
 and from the definitions
\begin{equation}\label{dm7}
  \dot{q}^*=\frac{d q}{d t^*},\hspace{1.00cm}
L^*=L(q,\dot{q}^*,t^*),\hspace{1.00cm}p^*=\frac{\partial
L^*}{\partial \dot{q}^*},
\end{equation}
the Lagragian equations can be obtained from
\begin{equation}\label{dm8}
\frac{d}{d t^*}(\frac{\partial L^*}{\partial
\dot{q}^*})-\frac{\partial L^*}{\partial q}=0,
\end{equation}
where $ H^*=\sum p^*\dot{q}^*-L^* $. Canonical commutation rule
and Schrodinger equation in this formalism are
\begin{equation}\label{dm9}
[q,p^*]=i,\hspace{2.00cm}H^*\psi=i\frac{\partial \psi}{\partial
t^*},
\end{equation}
but unfortunately uncertainty relations vanish as time goes to
infinity.\\
Perhaps one of the effective approaches in quantum mechanics of
dissipative systems is the idea of considering an environment
coupled to the main system and doing calculations for the total
system but at last for obtaining observables related to the main
system, the environment degrees of freedom must be eliminated.
The interested reader is referred to the Caldeira-Legget model
[7,8]. In this model the dissipative system is coupled with an
environment made by a collection of $ N $ harmonic oscillators
with masses $ m_n $ and frequencies $\omega_n $, the interaction
term in Hamiltonian is as follows
\begin{equation}\label{dm10}
H'=-q\sum_{n=1}^N c_n
x_n+q^2\sum_{n=1}^N\frac{c_n^2}{2m_n\omega_n^2},
\end{equation}
where  $ q $ and $ x_n $ denote coordinates of system and
environment respectively and the constants $ c_n $ are
called coupling constants.\\
The above coupling is not suitable for dissipative systems
containing a dissipation term proportional to velocity. In fact
with above coupling we can not obtain Heisenberg equation like $
\ddot{q}+\omega^2 q+\beta\dot{q}=\xi( t)$ for a damped harmonic
oscillator consistently and we can not study dissipative quantum
fields, for example, a dissipative vibrating medium with this
model. In this paper we generalize the Caldeira-Legget model to an
environment with continuous degrees of freedom by a coupling
similar to the coupling between a charged particle and the
electromagnetic field known as minimal coupling. In sections 2,
the idea of a minimal coupling is introduced and the quantum
dynamics of a damped vibrating string as the simplest scalar
field theory, is investigated. In section 3, quantum dynamics of
the string and it's environment is investigated. In section 4 some
transition probabilities indicating the way dissipation flows,
are obtained.
\section{Quantum dynamics of a damped vibrating string}
In this section we consider a damped vibrating string as the
dissipative system although the method is general and can be
applied to a general scalar field. Quantum mechanics of a damped
vibrating string with mass density $\lambda $, tention $ \mu $
 and lengh $ L $, can be investigated by
introducing a reservoir or an environment that interacts with the
string through a new kind of minimal coupling. Let the two ends
of the string be fixed in $ x=0 $ and $x=L $ respectively and
vibration be only in the $ y $ direction. If $ \psi(x,t) $ is the
wave function of the string, to quantizing $ \psi(x,t) $, we
assume $ \psi(x,t) $ to be a hermitian operator and can be
expanded in terms of orthogonal functions, $ \sin\frac{n\pi x}{L}$
as follows
\begin{equation} \label{d1}
\psi(x,t)=\sum_{n=1}^\infty\frac{1}{\sqrt{L\lambda\omega_n}}[a_n(t)+a_n^\dag(t)]\sin\frac{n\pi
x}{L},
\end{equation}
where $ \omega_n=\sqrt{\frac{\mu}{\lambda}}\frac{n\pi}{L} $ and $
a_n $ and $ a_n^\dag $ are annihilation and creation operator of
the string respectively and satisfy in any instante of time the
following commutation rule
\begin{equation}\label{d2}
[a_n(t),a_m^\dag(t)]=\delta_{nm},
\end{equation}
By definition of a conjugate canonical momentum density as
\begin{equation}\label{d3}
\pi_\psi(x,t)=i\sum_{n=1}^\infty\sqrt{\frac{\lambda\omega_n}{L}}
[a_n^\dag(t)-a_n(t)]\sin\frac{n\pi, x}{L}
\end{equation}
then from (\ref{d2}) $ \psi$, $\pi_\psi $ satisfy equal time
comutation relation
\begin{equation}\label{d4}
[\psi(x,t),\pi_\psi(x',t)]=i\delta(x-x').
\end{equation}
 Hamiltonian of string is defined by
\begin{equation}\label{d4.5}
H_s=\int_0^L d
x(\frac{\pi_\psi^2}{2\lambda}+\frac{1}{2}\mu\psi_x^2)=\sum_{n=1}^\infty\omega_n(a_n^\dag.
a_n+\frac{1}{2})
\end{equation}
 Let the total Hamiultonian, i.e., string plus environment be like
this
\begin{equation}\label{d5}
H(t)=\int_0^L d x
\frac{(\pi_\psi(x,t)-R(x,t))^2}{2\lambda}+\frac{1}{2}\mu\psi_x^2+H_B,
\end{equation}
where $ \psi_x $ denots derivative with respect to $ x$ and $\mu $
is a constant depending on string properties, $ H_B $ is the
reservoir Hamiltonian
\begin{equation}\label{d3}
H_B(t)=\sum_{n=1}^\infty\int_{-\infty}^{+\infty}d^3k
\omega_{\vec{k}} (b_{n\vec{k}}^\dag(t)
b_{n\vec{k}}(t)+\frac{1}{2}), \hspace{1.50 cm}
\omega_{\vec{k}}=|\vec{k}|.
\end{equation}
Annihilation and creation operators $ b_{n\vec{k}}$,
$b_{n\vec{k}}^\dag $, in any instant of time, satisfy the
following commutation relations
\begin{equation}\label{d4}
[b_{n\vec{k}}(t),b_{m\vec{k}'}^\dag(t)]=\delta_{nm}\delta(\vec{k}-\vec{k}'),
\end{equation}
and we will show later in section 3 that reservoir is an infinit
number of independent Klein-Gordon equations with a source term.
Operator $ R(x,t) $ have the basic role in interaction between
string and reservoir and is defined by
\begin{equation}\label{d5}
R(x,t)=\sum_{n=1}^\infty\int_{-\infty}^{+\infty}d^3k
[f(\omega_{\vec{k}})
b_{n\vec{k}}(t)+f^*(\omega_{\vec{k}})b_{n\vec{k}}^\dag(t)]\sin\frac{n\pi
x }{L},
\end{equation}
let us call the function $ f(\omega_k) $, the coupling function.
By using of (\ref{d4}), it can be shown easily that Heisenberg
equation for $ \psi(x,t) $ and $\pi_\psi(x,t)$  leads to
\begin{eqnarray}\label{d6}
&&\dot{\psi}(x,t)=i[H,\psi(x,t)]=\frac{\pi_\psi-R}{\lambda},\nonumber\\
&& \dot{\pi}_\psi(x,t)=i[H,\pi_\psi(x,t)]=\mu\psi_{xx},
\end{eqnarray}
which after omitting $\pi_\psi$, gives the following equation for
the damped vibrating string
\begin{equation}\label{d6.5}
\lambda\ddot{\psi}-\mu\psi_{xx}=-\dot{R}(x,t).
\end{equation}
 Using (\ref{d4}) the Heisenberg equation for $ b_{n\vec{k}}$, is
\begin{equation}\label{d7}
\dot{b}_{n\vec{k}}=i[H,b_{n\vec{k}}]=-i\omega_{\vec{k}}
b_{n\vec{k}}+if^*(\omega_{\vec{k}})\int_0^L
\dot{\psi}(x',t)\sin\frac{n\pi x'}{L}d x',
\end{equation}
with the following formal solution
\begin{equation}\label{d8}
b_{n\vec{k}}(t)=b_{n\vec{k}}(0)e^{-i\omega_{\vec{k}}
t}+if^*(\omega_{\vec{k}}) \int_0^t d t'
e^{-i\omega_{\vec{k}}(t-t')}\int_0^L
\dot{\psi}(x',t')\sin\frac{n\pi x'}{L}d x' ,
\end{equation}
substituting $ b_{n\vec{k}}(t) $ from (\ref{d8}) into
(\ref{d6.5}), using the relation $
\delta(x-x')=\frac{2}{L}\sum_{n=1}^\infty\sin\frac{n\pi
x}{L}\sin\frac{n\pi x'}{L} $ and at last integrating over $x'$
gives
\begin{eqnarray}\label{d9}
&&\lambda\ddot{\psi}-\mu\psi_{xx}+\int_0^t d
t'\dot{\psi}(x,t')\gamma(t-t')=\xi(x,t),\nonumber\\
&&\xi(x,t)=i\int_{-\infty}^{+\infty} d^3 k
\omega_{\vec{k}}[f(\omega_{\vec{k}})b_{n\vec{k}}(0)
e^{-i\omega_{\vec{k}}t}-f^*(\omega_{\vec{k}})b_{n\vec{k}}^\dag(0)e^{i\omega_{\vec{k}}t}]
\sin\frac{n\pi x}{L},\nonumber\\
&&\gamma(t)=4\pi L\int_0^\infty d\omega_{\vec{k}}
|f(\omega_{\vec{k}})|^2\omega_{\vec{k}}^3\cos\omega_{\vec{k}}t,
\end{eqnarray}
it is clear that the expectation value of $ \xi(x,t) $ in any
eigenstate of $ H_B $, is zero. For the following special choice
of coupling function
\begin{equation}\label{d10}
f(\omega_{\vec{k}})=\sqrt{\frac{\beta}{4\pi^2
L\omega_{\vec{k}}^3}},
\end{equation}
equation (\ref{d9}) takes the form
\begin{eqnarray}\label{d11}
&&\lambda\ddot{\psi}-\mu\psi_{xx}+\beta\dot{\psi}=\tilde{\xi}(x,t),\nonumber\\
&&\tilde{\xi}(x,t)=i\sqrt{\frac{\beta}{4\pi^2
L}}\int_{-\infty}^{+\infty} \frac{d^3 k}{\sqrt{\omega_{\vec{k}}}}
[b_{n\vec{k}}(0)e^{-i\omega_{\vec{k}}t}-
b_{n\vec{k}}^\dag(0)e^{i\omega_{\vec{k}}t}]\sin\frac{n\pi x}{L},
\end{eqnarray}
Hisenberg equation for $ a_n $ and $ a_n^\dag $ is
\begin{eqnarray}\label{d11.1}
&&\frac{\dot{a}_n+\dot{a}_n^\dag}{\sqrt{L\lambda\omega_n}}=i\sqrt{\frac{\omega_n}
{L\lambda}}(a_n^\dag-a_n)-\frac{1}{\lambda}\int_{-\infty}^{+\infty}
d^3k[f(\omega_{\vec{k}})b_{n\vec{k}}(t)+f^*(\omega_{\vec{k}})b_{n\vec{k}}^\dag(t)],\nonumber\\
&&i\sqrt{\frac{\lambda\omega_n}{L}}(\dot{a}_n^\dag-\dot{a}_n)=-\lambda\omega_n^2
\frac{(a_n^\dag+a_n)}{\sqrt{\lambda L\omega_n}}.
\end{eqnarray}
Definition $ A_n=\frac{a_n+a_n^\dag}{\sqrt{L\lambda\omega_n}} $
and $ B_n=i\sqrt{\frac{\lambda\omega_n}{L}}(a_n^\dag-a_n) $ and
using (\ref{d10}), we can easily obtain
\begin{eqnarray}\label{d11.2}
&&\ddot{A}_n+\omega_n^2A_n+\frac{\beta}{\lambda}\dot{A}_n=\zeta_n(t),\nonumber\\
&&\zeta_n(t)=i\sqrt{\frac{\beta}{4\pi^2\lambda^2L}}\int_{-\infty}^{+\infty}\frac{d^3k}{\sqrt{\omega_{\vec{k}}}}
[b_{n\vec{k}}(0)e^{-i\omega_{\vec{k}}t}-b_{n\vec{k}}^\dag(0)e^{i\omega_{\vec{k}}t}],
\end{eqnarray}
with the following solution
\begin{eqnarray}\label{d12}
&&A_n(t)=e^{-\frac{\beta t}{2\lambda}}(\hat{E}_ne^{i\Omega_n
t}+\hat{F}_n
e^{-i\Omega_n t})+M(t),\nonumber\\
&& M_n(t)=i\int_{-\infty}^{+\infty}d^3 k
\sqrt{\frac{\beta}{4\pi^2\lambda^2L\omega_{\vec{k
}}}}[\frac{b_{n\vec{k}}(0)}{\omega_n^2-\omega_{\vec{k}}^2-i\frac{\beta}{\lambda}
\omega_{\vec{k}}}e^{-i\omega_{\vec{k}}
t}-\frac{b_{n\vec{k}}^\dag(0)}{\omega_n^2-\omega_{\vec{k}}^2+\frac{i\beta}{\lambda}
\omega_{\vec{k}}}e^{i\omega_{\vec{k}}t}],\nonumber\\
&&
\end{eqnarray}
where $ \Omega_n=\sqrt{\omega_n^2-\frac{\beta^2}{4\lambda^2}}$.
Operators $ \hat{E}_n $ and $ \hat{F}_n $, are specified by
initial conditions
\begin{eqnarray}\label{d13}
\hat{E}_n+\hat{F}_n&=&A_n(0)-M_n(0),\nonumber\\
(\frac{-\beta}{2\lambda}+i\Omega_n)\hat{E}_n-
(\frac{\beta}{2\lambda}+i\Omega_n)\hat{F}_n&=&
\dot{A}_n(0)-\dot{M}_n(0),
\end{eqnarray}
solving above equations and substituting $ \hat{E}_n$ and $
\hat{F}_n$ in (\ref{d12}) one obtains
\begin{eqnarray}\label{d14}
&&A_n(t)=e^{-\frac{\beta t}{2\lambda}} \{A_n(0)\cos\Omega_n t+
\frac{\beta}{2\lambda\Omega_n}A_n(0)\sin\Omega_n t  -\frac{\beta
M_n(0)}{2\lambda\Omega_n}\sin\Omega_nt\nonumber\\
&&-M_n(0)\cos\Omega_n t
+\frac{\dot{A}_n(0)-\dot{M}_n(0)}{\Omega_n}\sin\Omega_n
t\}+M_n(t).
\end{eqnarray}
It is clear from (\ref{d11.1}) that $ \dot{A}_n(0) $ is dependent
on string and reservoir operators in $ t=0 $ . Substituting $
\dot{A}_n(t) $ from (\ref{d14}) in (\ref{d8}) we can find a stable
solution for $ b_{n\vec{k}}(t) $ in $ t\rightarrow\infty $ as
\begin{eqnarray}\label{d15}
&& b_{n\vec{k}}(t)=b_{n\vec{k}}(0)e^{-i\omega_{\vec{k}}
t}-i\sqrt{\frac{L\beta }{16\pi^2
\omega_{\vec{k}}^3}}\frac{e^{-i\omega_{\vec{k}}
t}}{(\omega_n^2-\omega_{\vec{k}}^2-\frac{i\beta}{\lambda}\omega_{\vec{k}})}\{\omega_n^2
A_n(0)-\omega_n^2M_n(0)\nonumber\\
&&+i\omega_{\vec{k}}(\dot{A}_n(0)-\dot{M}_n(0))\}\nonumber\\
&&+\frac{i\beta}{8\pi^2\lambda\sqrt{\omega_{\vec{k}}^3}}
\int_{-\infty}^{+\infty} d^3
k'\sqrt{\omega_{\vec{k'}}}\{\frac{b_{n\vec{k'}}(0)}{\omega_n^2-\omega_{\vec{k'}}^2-\frac{i\beta}{\lambda}\omega_{\vec{k'}}}
\frac{\sin\frac{(\omega_{\vec{k}}-\omega_{\vec{k'}}t)
}{2}t}{\frac{(\omega_{\vec{k}}-\omega_{\vec{k'}})}{2}}e^{\frac{-i(\omega_{\vec{k}}+\omega_{\vec{k'}})t}{2}}\nonumber\\
&&+\frac{b_{n\vec{k'}}^\dag(0)}{\omega^2-\omega_{\vec{k'}}^2+\frac{i\beta}{m}\omega_{\vec{k'}}}
\frac{\sin\frac{(\omega_{\vec{k}}+
\omega_{\vec{k'}})}{2}t}{\frac{(\omega_{\vec{k}}+\omega_{\vec{k'}})}{2}}
e^{\frac{i(\omega_{\vec{k'}}-\omega_{\vec{k}})t}{2}}\},
\end{eqnarray}
now substituting $ b_{n\vec{k}}(t) $ from (\ref{d15}) in
(\ref{d11.1}) and using (\ref{d14}), one can show that
\begin{equation}\label{d15.1}
B_n(t)=\lambda\dot{A}_n(t)+\sum_{n=1}^\infty\int_{-\infty}^{+\infty}
d^3k\sqrt{\frac{\beta}{4\pi^2L\omega_{\vec{k}}^3}}[b_{n\vec{k}}(t)+b_{n\vec{k}}^\dag(t)).
\end{equation}
A vector in fock space of string can be written like this
\begin{equation}\label{d15.2}
|\Phi\rangle_s=\sum_{j=0}^\infty\sum_{n_1,...,n_j=1}^\infty\Phi_{n_1,...,n_j}|n_1\rangle_s\otimes...\otimes|n_j\rangle_s
\end{equation}
where $ |n\rangle_s $ denotes the state of a single particle in
mode $n$, with corresponding wave function $\langle x|n\rangle=
\sqrt{\frac{2}{L}}\sin\frac{n\pi x}{L} $. Operators $ a_n(0) $
and $ a_n^\dag(0) $ act on basis vectors $
|n_1\rangle_s\otimes...\otimes|n_j\rangle_s\equiv|n_1,...,n_j\rangle_s
$ as follows
\begin{eqnarray}\label{d15.3}
&&a_m(0)|n_1,...,n_j\rangle_s=\sum_{r=1}^j\delta_{n_r,m}|n_1,...,n_{r-1},n_{r+1},...,n_j
\rangle_s,\nonumber\\
&&a_m^\dag(0)|n_1,...,n_j\rangle_s=|m,n_1,...,n_j\rangle_s,
\end{eqnarray}
also a vector in fock space of reservoir can be written as
\begin{equation}\label{d15.2}
|\Psi\rangle_B=\sum_{j=0}^\infty\sum_{\nu_1,...,\nu_j=1}^\infty\int
d^3k_1...d^3k_j\Psi_{n_1,...,n_j}(\vec{k}_1,...,\vec{k}_j)|\vec{k}_1,\nu_1\rangle_B
\otimes...\otimes|\vec{k}_j,\nu_j\rangle_B.
\end{equation}
In subsequent section we show that reservoir is infinit number of
independent Klein-Gordon fields and we can interpret $
|\vec{k},\nu\rangle_B $ as a single particle state belong to $ \nu
$th Klein-Gordon field with corresponding momentum, $ \vec{k} $.
 Operators $ b_{n\vec{q}}(0) $ and $ b^{\dag}_{n\vec{q}}(0) $ act on basis
vectors
$|\vec{k}_1,\nu_1\rangle_B\otimes...\otimes|\vec{k}_j,\nu_j\rangle_B\equiv|\vec{k}_1,
\nu_1,...,\vec{k}_j,\nu_j\rangle_B
$ as
\begin{eqnarray}\label{d15.3}
&&b_{n\vec{q}}(0)|\vec{k}_1,\nu_1,...,\vec{k}_j,\nu_j\rangle_B\nonumber\\
&&=\sum_{r=1}^j\delta_{n,\nu_r}\delta(\vec{q}-\vec{k}_r)|\vec{k}_1,\nu_1,...\vec{k}_{r-1},
\nu_{r-1},\vec{k}_{r+1},\nu_{r+1},...,\vec{k}_j,\nu_j\rangle_B,\nonumber\\
&&b^{\dag}_{n\vec{q}}(0)|\vec{k}_1,\nu_1,...,\vec{k}_j,\nu_j\rangle_B=|\vec{q},n,\vec{k}_1,
\nu_1,...,\vec{k}_j,\nu_j\rangle_B.
\end{eqnarray}
If the state of system in $ t=0 $ is taken to be $
|\psi(0)\rangle=|0\rangle_B\otimes|m_1,...m_r\rangle_s  $ where $
|0\rangle_B $ is vacuum state of reservoir and
$|m_1,...m_r\rangle_s $ an excited state of the Hamiltonian $ H_s
$ then by making use of (\ref{d14}), (\ref{d15}) and
(\ref{d15.1}) it can be shown that
\begin{eqnarray}\label{d16}
&& lim_{t\rightarrow\infty}[ _B\langle0|\otimes
\hspace{00.20cm}_s\langle m_1,...,m_r|: \int_0^L dx[
\frac{1}{2}\lambda\dot{\psi}^2+\frac{1}{2}\mu \psi_x^2]
:|m_1,...,m_r\rangle_s\otimes|0\rangle_B]= 0,\nonumber\\
&&lim_{t\rightarrow\infty}[ _B\langle0|\otimes
\hspace{00.20cm}_s\langle m_1,...,m_r|: \sum_{n=1}^\infty\omega_n
a_n^\dag(t) a_n(t)
:|m_1,...,m_r\rangle_s\otimes|0\rangle_B]\nonumber\\
&&=lim_{t\rightarrow\infty}\{\sum_{n=1}^\infty\frac{L\beta^2\omega_n^4}{16\pi^2
\lambda}\hspace{00.25 cm}|\int_{-\infty}^{+\infty}\frac{d
x}{x}\frac{e^{ixt}}{\omega^2-x^2+i\frac{\beta}{\lambda}x}|^2\nonumber\\
&&\times_ s\langle m_r,...m_1| : A_n^2(0) : |m_1,...m_r
\rangle_s\}\simeq\frac{\beta^2}{8\lambda^2}\sum_{i=1}^r\frac{1}{\omega_{m_i}}.
\end{eqnarray}
where $ :\hspace{0.4cm }: $ denotes the normal ordering operator.
Now by substituting $ b_{n\vec{k}}(t) $ from (\ref{d15}) into
(\ref{d3}), we have
\begin{eqnarray}\label{d17}
&&lim_{t\rightarrow\infty}[ _B\langle0|\otimes\hspace{00.10cm}
_s\langle m_r,...,m_1|:\int d^3k\sum_{n=1}^\infty \omega_{\vec{k}} b_{n\vec{k}}^\dag b_{n\vec{k}}:|m_1,...m_r\rangle_s\otimes|0\rangle_B]\nonumber\\
&&=\frac{\beta}{2\pi\lambda}\sum_{i=1}^r
\omega_{m_i}^3\int_0^\infty \frac{d
x}{(\omega_{m_i}^2-x^2)^2+\frac{\beta^2}{\lambda^2}x^2},\nonumber\\
&&+\frac{\beta}{2\pi \lambda
}\sum_{i=1}^r\omega_{m_i}\int_0^\infty \frac{x^2d x
}{(\omega_{m_i}^2-x^2)^2+\frac{\beta^2}{\lambda^2}x^2}.
\end{eqnarray}
\section{Quantum field of reservoir}
Let us define the operators $ Y_n(\vec{x},t) $ and $
\Pi_n(\vec{x},t) $ as follows
\begin{eqnarray}\label{d18}
&&Y_n(\vec{x},t)=\int_{-\infty}^{+\infty} \frac{d^3
k}{\sqrt{2(2\pi)^3\omega_{\vec{k}}}}(
b_{n\vec{k}}(t)e^{i\vec{k}.\vec{x}}+b_{n\vec{k}}^\dag(t)e^{-i\vec{k}.\vec{x}}),\nonumber\\
&&\Pi_n(\vec{x},t)=i\int_{-\infty}^{+\infty} d^3 k
\sqrt{\frac{\omega_{\vec{k}}}{2(2\pi)^3}}( b_{n\vec{k}}^\dag
(t)e^{-i\vec{k}.\vec{x}}-b_{n\vec{k}}(t)e^{i\vec{k}.\vec{x}}),
\end{eqnarray}
then using commutation relations (\ref{d4}), one can show that $
Y_n(\vec{x},t) $ and $ \Pi_m (\vec{x},t)$, satisfy the equal time
commutation relations
\begin{equation}\label{d19}
[Y_n(\vec{x},t),\Pi_m(\vec{x'},t)]=i\delta_{nm}\delta(\vec{x}-\vec{x'}),
\end{equation}
furthermore by substituting $ b_{n\vec{k}}(t)$ from ( \ref{d8}) in
(\ref{d18}) we obtain
\begin{eqnarray}\label{d20}
&&\frac{\partial\Pi_n(\vec{x},t)}{\partial
t}=\nabla^2Y_n+L\dot{A_n}(t) P(\vec{x}),\hspace{0.5 cm
}P(\vec{x})=Re{ \int_{-\infty}^{+\infty}
d^3 k\sqrt{\frac{\omega_{\vec{k}}}{2(2\pi)^3}}f(\omega_{\vec{k}})e^{-i\vec{k}.\vec{x}}},\nonumber\\
&&\Pi_n(\vec{x},t)=\frac{\partial Y_n}{\partial t}-L\dot{A_n}(t)
Q(\vec{x}),\hspace{0.5 cm} Q(\vec{x})= Im{\int_{-\infty}^{+\infty}
d^3
k\frac{f(\omega_{\vec{k}})}{\sqrt{2(2\pi)^3\omega_{\vec{k}}}}e^{-i\vec{k}.\vec{x}}},\nonumber\\
&&
\end{eqnarray}
so for any $ n $, $ Y_n(\vec{x},t)$ satisfies the following
source included Klein-Gordon equation
\begin{equation}\label{d21}
\frac{\partial^2Y_n}{\partial
t^2}-\nabla^2Y_n=L\ddot{A_n}(t)Q(\vec{x})+2\dot{A_n}(t)P(\vec{x}),
\end{equation}
with the corresponding Lagrangian density as follows
\begin{equation}\label{d22}
\pounds_n=\frac{1}{2}(\frac{\partial Y_n}{\partial
t})^2-\frac{1}{2}\vec{\nabla Y_n}.\vec{\nabla
Y_n}-L\dot{A_n}Q(\vec{x})\frac{\partial Y_n}{\partial
t}+L\dot{A_n}P(\vec{x})Y_n.
\end{equation}
It is clear that the reservoir is made by an infinit number of
massless Klein-Gordon fields containing the source term $
2\ddot{A_n}Q(\vec{x})+2\dot{A_n}P(\vec{x})$. Hamiltonian density
for (\ref{d21}) is
\begin{equation}\label{d22.5}
\aleph_n=\frac{(\Pi_n+L\dot{A_n}Q)^2}{2}+\frac{1}{2}|\vec{\nabla
Y_n}|^2-L\dot{A_n} P(\vec{x,})Y_n,
\end{equation}
and equations (\ref{d20}) are Heisenberg equations for fields $
Y_n $ and $ \Pi_n$. If we obtain $ b_{n\vec{k}} $ and $
b_{n\vec{k}}^\dag $ from (\ref{d18}) in terms of $ Y_n $ and $
\Pi_n $  and substitute them in (\ref{d3}), we find
\begin{equation}\label{d22.75}
H_B=\int_{-\infty}^{+\infty}d^3k
\omega_{\vec{k}}(b_{n\vec{k}}^\dag
b_{n\vec{k}}+\frac{1}{2})=\frac{\Pi_n^2}{2}+\frac{1}{2}|\vec{\nabla
Y_n }|^2.
\end{equation}
\section{Transition probabilities}
We can write the Hamiltonian (\ref{d5}) as
\begin{eqnarray}\label{d23}
&&H=H_0+H',\nonumber\\
&&H_0=H_s+H_B=\sum_{n=1}^\infty(a_n^\dag a_n+\frac{1}{2})\omega_n+
\sum_{n=1}^\infty\int_{-\infty}^\infty d^3k \omega_{\vec{k}}(b_{n\vec{k}}^\dag b_{n\vec{k}}+\frac{1}{2}) \nonumber\\
&&H'=-\int_0^L
dx\frac{\pi_\psi(x,t)}{\lambda}R(x,t)+\frac{R^2(x,t)}{2\lambda},
\end{eqnarray}
and in interaction picture we can write
\begin{eqnarray}\label{d24}
&& a_{nI}(t)=e^{iH_0 t}a_n(0)e^{-iH_0 t}=a_n(0)e^{-i\omega_n t},\nonumber\\
&& b_{n\vec{k}I}(t)=e^{iH_0 t}b_{n\vec{k}}(0)e^{-iH_0
t}=b_{n\vec{k}}(0)e^{-i\omega_{\vec{k}} t},
\end{eqnarray}
terms $ \frac{R}{\lambda} \pi_\psi $ and $ \frac{R^2}{2\lambda} $
are of the first order and second order of damping respectively,
therefore for sufficiently weak damping, $ \frac{R^2}{2\lambda}$
is small in comparison with $ \frac{R}{\lambda} \pi_\psi $.
Furthermore  $ \frac{R^2}{2\lambda} $ has not any role in those
transition probabilities where initial and final states of
Hamiltonian of vibrating string are different, hence we can
neglect the term $\frac{R^2}{2\lambda} $ and estimate $ H' $ by $
-\frac{ R}{\lambda} \pi_\psi $. Substituting $ a_{nI} $ and $
b_{n\vec{k}I} $ from (\ref{d24}) into  $ -\frac{ R}{\lambda}
\pi_\psi $, one obtains $ H_I' $ in interaction picture, as
\begin{eqnarray}\label{d25}
&&H_I'=-\frac{iL}{2\lambda}\sum_{n=1}^\infty\sqrt{\frac{\lambda\omega_n}{L}}\int_{-\infty}^{+\infty}d^3
k (f(\omega_{\vec{k}}) a_n^\dag(0) b_{n\vec{k}}(0)
e^{i(\omega_n-\omega_{\vec{k}})t}\nonumber\\
&&+f^*(\omega_{\vec{k}}) a_n^\dag(0) b_{n\vec{k}}^\dag(0)
e^{i(\omega_n+\omega_{\vec{k}} )t} -f(\omega_k) a_n(0)
b_{n\vec{k}}(0) e^{-i(\omega_{\vec{k}}
+\omega_n)t}\nonumber\\
&&-f^*(\omega_{\vec{k}})a_n(0) b_{n\vec{k}}^\dag(0)
e^{i(\omega_{\vec{k}}-\omega_n) t}),
\end{eqnarray}
 terms  containing just $
a_n(0) b_{n\vec{k}}(0) $ and $ a_n^\dag(0) b_{n\vec{k}}^\dag(0) $
violate the conservation of energy in the first order
perturbation, because $ a_n(0) b_{n\vec{k}}(0) $ destroys an
excited state of string while at the same time destroying a
reservoir excitation state and $ a_n^\dag(0) b_{n\vec{k}}^\dag(0)
$ creates an excited state of vibrating string, while creating an
excited reservoir state at the same time, therefore we neglect
the terms involving $ a_n(0) b_{n\vec{k}}(0) $ and $ a_n^\dag(0)
b_{n\vec{k}}^\dag(0) $, because of energy conservation and
estimate $ H'_I $ by
\begin{eqnarray}\label{d26}
&&H_I'=-\frac{i}{2}\sqrt{\frac{L}{\lambda}}\sum_{n=1}^\infty\int_{-\infty}^{+\infty}d^3
k \sqrt{\omega_n}[f(\omega_{\vec{k}})a_n^\dag(0) b_{n\vec{k}}(0)
e^{i(\omega_n-\omega_{\vec{k}})t}\nonumber\\
&&-f^*(\omega_{\vec{k}}) a_n(0) b_{n\vec{k}}^\dag(0)
e^{-i(\omega_n-\omega_{\vec{k}})t}].
\end{eqnarray}
The time evolution of density operator in interaction picture is
as follows [10]
\begin{equation}\label{d27}
\rho_I(t)=U_I(t,t_0)\rho_I(t_0)U_I^\dag(t,t_0),
\end{equation}
where $ U_I $ is the time evolution operator, which in first
order perturbation is
\begin{eqnarray}\label{d28}
 &&U_I(t,t_0=0)=1-i \int_0^t d t_1 H'_I(t_1)=\nonumber\\
 && 1-\frac{1}{2}\sqrt{\frac{L}{\lambda}}\int_{-\infty}^{+\infty}d^3
 k\sum_{n=1}^\infty\sqrt{\omega_n}
[f(\omega_{\vec{k}}) a_n^\dag(0) b_{n\vec{k}}(0)
e^{\frac{i(\omega_n-\omega_{\vec{k}})
t}{2}}\nonumber\\
&&-f^*(\omega_{\vec{k}}) a_n(0) b_{n\vec{k}}^\dag(0)
e^{\frac{-i(\omega_n-\omega_{\vec{k}})t}{2}}]
\frac{\sin\frac{(\omega_n-\omega_{\vec{k}})}{2}t}{\frac{(\omega_n-\omega_{\vec{k}})}{2}}.
\end{eqnarray}
Let $ \rho_I(0)=|m,...,m\rangle^{r}_s\hspace{0.20 cm}_s^{r}\langle
m,...,m|\otimes|0\rangle_B\hspace{0.20 cm}_B \langle
 0|$ where $ |0\rangle_B $ is the vacuum state of reservoir and $ |m,...,m\rangle^{r}_s $ is an
 excited state of vibrating string, from now on by $|m,...,m\rangle^{r}_s$, we mean a string
 state containing $r$ phonons of mode $m$, substituting $ U_I(t,0) $ from
 (\ref{d28}) in (\ref{d27}) and taking trace  over reservoir parameters, we obtain
 \begin{eqnarray}\label{d28.1}
 &&\rho_{sI}(t):=Tr_B(\rho_I(t))=|m,...,m\rangle^{r}_s\hspace{0.20
cm}_s^{r}\langle m,...,m|\nonumber\\
&&+\frac{L\omega_m}{4\lambda}|m,...,m\rangle^{r-1}_s\hspace{0.20
cm}_s^{r-1}\langle m,...,m|\int_{-\infty}^{+\infty} d^3 p
|f(\omega_{\vec{p}})|^2
\frac{\sin^2\frac{(\omega_{\vec{p}}-\omega_m)}{2}t}
{(\frac{\omega_{\vec{p}}-\omega_m}{2})^2},\nonumber\\
&&
\end{eqnarray}
where we have used the formula $ Tr_B [ |\vec{k},n\rangle_B
\hspace{00.20cm}_B\langle\vec{k'},s|
]=\delta_{ns}\delta(\vec{k}-\vec{k'})$. In large time
approximation, we can write
$\frac{\sin^2\frac{(\omega_{\vec{p}}-\omega_m)}{2}t}{(\frac{\omega_{\vec{p}}-\omega_m}{2})^2}=2\pi
t\delta(\omega_{\vec{p}}-\omega_m)$, which leads to the following
relation for density matrix
\begin{equation}\label{d28.2}
\rho_{sI}(t)=|m,...,m\rangle^{r}_s\hspace{0.20 cm}_s^{r}\langle
m,...,m|+\frac{2L\pi^2\omega_m^3
t|f(\omega_m)|^2}{\lambda}|m,...,m\rangle^{r-1}_s\hspace{0.20
cm}_s^{r-1}\langle m,...,m|,
\end{equation}
from density matrix we can calculate the probability of transition
$|m,...,m\rangle^{r}_s\rightarrow|m,...,m\rangle^{r-1}_s $ as
\begin{eqnarray}\label{d29}
&&\Gamma_{|m,...,m\rangle^{r}_s\rightarrow
|m,...,m\rangle^{r-1}_s}=Tr [|m,...,m\rangle^{r-1}_s\hspace{0.20
cm}_s^{r-1}\langle m,...,m|\rho(t)]=\nonumber\\
&&Tr_s[|m,...,m\rangle^{r-1}_s\hspace{0.20 cm}_s^{r-1}\langle
m,...,m|\rho_{sI}(t)]=\frac{2L\pi^2 \omega_m^3  t
|f(\omega_m)|^2}{\lambda},
\end{eqnarray}
where $ Tr_s $ means taking trace over string eigenstates. For
the special choice (\ref{d10}), above transition probability
becomes
\begin{equation}\label{d30}
\Gamma_{|m,...,m\rangle^{r}_s\rightarrow
|m,...,m\rangle^{r-1}_s}=\frac{\beta t}{2\lambda},
\end{equation}
which shows that the rate of phonon number reduction (energy
flow), is constant. Now consider the case where the reservoir is
an excited state in $ t=0$ for example $
\rho_I(0)=|m,...,m\rangle^{r}_{s^{r}}\hspace{0.20 cm}_s^{r}\langle
m,...,m|\otimes|\vec{p}_1,\nu_1,...\vec{p}_j,\nu_j\rangle_{B\hspace{0.20
cm}B} \langle
 \vec{p}_1,\nu_1,...\vec{p}_j,\nu_j| $ where
 $ |\vec{p}_1,\nu_1...\vec{p}_j,\nu_j\rangle_B $ denotes a reservoir state
  containing $ j $ quanta with  momenta $ \vec{p}_1,...,\vec{p}_j$
 belonging to the $\nu_1$,$\nu_2$,...,$\nu_j$th,
 Klein-Gordon field, respectively then by making use of
\begin{eqnarray}\label{d31}
 Tr_B[ b_{n\vec{k}}^\dag |\vec{p}_1,\nu_1...\vec{p}_j,\nu_j\rangle_{B\hspace{0.20 cm}B} \langle
 \vec{p}_1,\nu_1...\vec{p}_j,\nu_j| b_{m\vec{k'}}]&=&\delta_{nm}\delta(\vec{k}-\vec{k'}),\nonumber\\
 Tr_B[b_{n\vec{k}} |\vec{p}_1,\nu_1...\vec{p}_j,\nu_j\rangle_{B\hspace{0.20 cm}B} \langle
 \vec{p}_1,\nu_1...\vec{p}_j,\nu_j|b_{m\vec{k'}}^\dag]&=&\sum_{l=1}^j
\delta_{n,\nu_l}\delta_{m,\nu_l}\delta(\vec{k}-\vec{p}_l)\delta(\vec{k'}-\vec{p}_l),
 \end{eqnarray}
  we find
 \begin{eqnarray}\label{d32}
 \rho_{sI}(t)&=&|m,...,m\rangle^{r}_s\hspace{0.20 cm}_s^{r}\langle
m,...,m|+\nonumber\\
&+&\frac{L}{4\lambda}\sum_{r=1}^j\omega_{\nu_r}|\nu_r,m,...,m\rangle^{r}_s\hspace{0.20
cm}_s^{r}\langle \nu_r,m,...,m| |f(\omega_{\vec{p}_r})|^2
\frac{\sin^2 \frac{(\omega_{\vec{p}_r}-\omega_{\nu_r})}{2}t}{(
\frac{\omega_{\vec{p}_l}-\omega_{\nu_r}}{2})^2}\nonumber\\
&+&\frac{L}{4\lambda}|m,...,m\rangle^{r-1}_\omega\hspace{0.20
cm}_\omega^{r-1}\langle m,...,m|\int_{-\infty}^{+\infty} d^3 k
|f(\omega_{\vec{k}})|^2
\frac{\sin^2\frac{(\omega_{\vec{k}}-\omega_m)}{2}t}{(
\frac{\omega_{\vec{k}}-\omega_m}{2})^2},
\end{eqnarray}
which gives the transition probability for $ |m,...,m\rangle^{r}_s
\rightarrow|m,...,m\rangle^{r-1}_s $ and
$|m,...,m\rangle^{r}_s\rightarrow |\nu,m,...,m\rangle^{r}_s $,
respectively, as follows
\begin{eqnarray}\label{d32}
\Gamma_{|m,...,m\rangle^{r}_s \rightarrow
|m,...,m\rangle^{r-1}_s}&=&Tr_s[|m,...,m\rangle^{r-1}_s\hspace{0.20
cm}_s^{r-1}\langle m,...,m| \rho_{sI}(t)]=\frac{2L\pi^2
\omega_m^3 t}{\lambda}|f(\omega_m)|^2,\nonumber\\
\Gamma_{|m,...,m\rangle^{r}_{s}\rightarrow
|\nu,m,...,m\rangle^{r}_s}&=&Tr_s[|\nu,m,...,m\rangle^{r}_s\hspace{0.30
cm}_s^{r}
\langle |\nu,m,...,m| \rho_{sI}(t)]\nonumber\\
&=&\frac{\pi L\omega_\nu t }{2\lambda}
\sum_{r=1}^j|f(\omega_{\vec{p}_r})|^2 \delta(
\omega_{\vec{p}_r}-\omega_\nu).
\end{eqnarray}
For the choice (\ref{d10}), we find
\begin{eqnarray}\label{d33}
&&\Gamma_{|m,...,m\rangle^{r}_s\rightarrow
|m,...,m\rangle^{r-1}_s}=\frac{\beta
t}{2\lambda},\nonumber\\
&&\Gamma_{|m,...,m\rangle^{r}_s\rightarrow|\nu,m,...,m\rangle^{r}_s}=\frac{\beta\omega_\nu
t}{8\pi \lambda } \sum_{r=1}^j
\frac{1}{\omega_{\vec{p}_r}}^3\delta(
\omega_{\vec{p}_r}-\omega_\nu).
\end{eqnarray}
Another important case is when the reservoir has a
Maxwell-Boltzman distribution, so let
$\rho_I(0)=|m,...,m\rangle^{r}_s\hspace{0.20 cm}_s^{r}\langle
m,...,m| \otimes \rho_B^T $ where\\  $
\rho_B^T=\frac{e^{\frac{-H_B}{K T}}}{TR_B(e^{\frac{-H_B}{KT}})}
$, then by making use of following relations
\begin{eqnarray}\label{d34}
&&Tr_B[ b_{n\vec{k}}\rho_B^T b_{m\vec{k'}}]=Tr_B[
b_{n\vec{k}}^\dag \rho_B^T
b_{m\vec{k'}}^\dag]=0,\nonumber\\
&& Tr_B[b_{n\vec{k}}\rho_B^T
b_{m\vec{k'}}^\dag]=\delta_{nm}\delta(\vec{k}-\vec{k'})
e^{-\frac{\omega_{\vec{k}}}{K T}},\nonumber\\
&&Tr_B[ b_{n\vec{k}}^\dag \rho_B^T
b_{m\vec{k'}}]=\delta_{nm}\delta(\vec{k}-\vec{k'}),
\end{eqnarray}
we can obtain the density operator $ \rho_{sI}(t) $ in interaction
picture as
 \begin{eqnarray}\label{d35}
 &&\rho_{sI}(t):=Tr_B [\rho_I(t)]=|m,...,m\rangle^{r}_s\hspace{0.20 cm}_s^{r}\langle
m,...,m|\nonumber\\
&&+\frac{L}{4\lambda}\sum_{n=1}^\infty
\omega_n|n,m,...,m\rangle^{r}_s\hspace{0.20 cm}_s^{r}\langle
n,m,...,m| \int_{-\infty}^{+\infty} d^3 k |f(\omega_{\vec{k}})|^2
\frac{\sin^2 \frac{(\omega_{\vec{k}}-\omega_n)}{2}t}{(
\frac{\omega_{\vec{k}}-\omega_n}{2})^2}e^{-\frac{\omega_{\vec{k}}}{K T}}\nonumber\\
&&+\frac{L\omega_m}{4\lambda}|m,...,m\rangle^{r-1}_s\hspace{0.20
cm}_s^{r-1}\langle m,...,m|\int_{-\infty}^{+\infty} d^3 k
|f(\omega_{\vec{k}})|^2 \frac{\sin^2
\frac{(\omega_{\vec{k}}-\omega_m)}{2}t}{(
\frac{\omega_{\vec{k}}-\omega_m}{2})^2},
\end{eqnarray}
which accordingly gives the following transition probabilities in
long time approximation
\begin{eqnarray}\label{d35}
&&\Gamma_{|m,...,m\rangle^{r}_s\rightarrow
|m,...,m\rangle^{r-1}_s}=Tr_s[|m,...,m\rangle^{r-1}_s\hspace{0.20
cm}_s^{r-1}\langle m,...,m| \rho_{sI}(t)]=\frac{2L\pi^2\omega_m^3
t}{\lambda}|f(\omega_m)|^2,\nonumber\\
&&\Gamma_{|m,...,m\rangle^{r}_s\rightarrow
|\nu,m,...,m\rangle^{r}_s}=Tr_s[|\nu,m,...,m\rangle^{r}_s\hspace{0.20
cm}_s^{r}\langle \nu,m,...,m|
\rho_{sI}(t)]=\nonumber\\
&&=\frac{2L\pi^2\omega_m^3
t}{\lambda}|f(\omega_m)|^2e^{-\frac{\omega_m}{K T}},
\end{eqnarray}
substituting (\ref{d10}) in these recent relations, we find
\begin{eqnarray}\label{d36}
&&\Gamma_{|m,...,m\rangle^{r}_s\rightarrow
|m,...,m\rangle^{r-1}_s}=\frac{\beta
t}{2\lambda},\nonumber\\
&&\Gamma_{|m,...,m\rangle^{r}_s\rightarrow
|\nu,m,...,m\rangle^{r}_s}=\frac{\beta
t}{2\lambda}e^{-\frac{\omega}{K T}}.
\end{eqnarray}
 So in very low temperatures the energy flows from oscillator to the reservoir by the rate
  $ \frac{\beta}{2\lambda}$ and no energy flows from
  reservoir to oscillator.
 \section{Concluding remarks}
 The Caldeira-Legget model generalized to the case where the environment
 has continuous degrees of freedom, for example, a Klein-Gordon field or an infinit number of
  Klein-Gordon fields. A minimal coupling method introduced which leads to a consistent
  investigation of the quantum dynamics of a large class of quantum dissipative
  systems. By choosing different coupling functions in
 (\ref{d5}), we could investigate another forms of dissipation.
 The rate of energy dissipation (energy flowing between the system
 and it's environment), was a constant. This problem can be extended to
 the case where the field $R$, becomes a general field for example a vector field, which is
 suitable for investigating three-dimensional quantum dissipative models.

\end{document}